\documentclass[preprint2]{aastex62}
\usepackage{amsmath}
\usepackage{color}
\begin{document}

\title{Constraining Exoplanet Metallicities and Aerosols with ARIEL:\\
An Independent Study by the Contribution to ARIEL Spectroscopy of Exoplanets (CASE) Team}

\email{rzellem@jpl.nasa.gov}
\author{Robert T. Zellem}
\affil{Jet Propulsion Laboratory, California Institute of Technology, 4800 Oak Grove Drive, Pasadena, California 91109, USA}

\author{Mark R. Swain}
\affil{Jet Propulsion Laboratory, California Institute of Technology, 4800 Oak Grove Drive, Pasadena, California 91109, USA}

\author{Nicolas B. Cowan}
\affil{Department of Physics, McGill University, 3600 rue University, Montr\'{e}al, Quebec, H3A 2T8, Canada}
\affil{Department of Earth \& Planetary Sciences, McGill University, 3450 rue University, Montr\'{e}al, Quebec, H3A 0E8, Canada}

\author{Geoffrey Bryden}
\affil{Jet Propulsion Laboratory, California Institute of Technology, 4800 Oak Grove Drive, Pasadena, California 91109, USA}

\author{Thaddeus D. Komacek}
\affil{Department of Planetary Sciences, Lunar and Planetary Laboratory, University of Arizona, 1629 East University Boulevard, University of Arizona, Tucson, AZ 85721, USA}
\affil{Department of the Geophysical Sciences, University of Chicago, 5734 S. Ellis Avenue Chicago, Illinois 60637, USA}

\author{Mark Colavita}
\affil{Jet Propulsion Laboratory, California Institute of Technology, 4800 Oak Grove Drive, Pasadena, California 91109, USA}

\author{David Ardila}
\affil{Jet Propulsion Laboratory, California Institute of Technology, 4800 Oak Grove Drive, Pasadena, California 91109, USA}

\author{Gael M. Roudier}
\affil{Jet Propulsion Laboratory, California Institute of Technology, 4800 Oak Grove Drive, Pasadena, California 91109, USA}

\author{Jonathan J. Fortney}
\affil{Department of Astronomy \& Astrophysics, University of California, Santa Cruz, CA, USA}

\author{Jacob Bean}
\affil{Department of Astronomy and Astrophysics, University of Chicago, 5640 South Ellis Avenue, Chicago, IL 60637, USA}

\author{Michael R. Line}
\affil{School of Earth \& Space Exploration, Arizona State University, Tempe, AZ 85281, USA}

\author{Caitlin A. Griffith}
\affil{Department of Planetary Sciences, Lunar and Planetary Laboratory, University of Arizona, 1629 East University Boulevard, University of Arizona, Tucson, AZ 85721, USA}

\author{Evgenya L. Shkolnik}
\affil{School of Earth \& Space Exploration, Arizona State University, Tempe, AZ 85281, USA}

\author{Laura Kreidberg}
\affil{Harvard Society of Fellows, 78 Mt. Auburn St., Cambridge, MA 02138, USA}
\affil{Harvard-Smithsonian Center for Astrophysics, 60 Garden St., Cambridge, MA 02138}

\author{Julianne I. Moses}
\affil{Space Science Institute, 4750 Walnut Street, Suite 205, Boulder, CO, 80301, USA}

\author{Adam P. Showman}
\affil{Department of Planetary Sciences, Lunar and Planetary Laboratory, University of Arizona, 1629 East University Boulevard, University of Arizona, Tucson, AZ 85721, USA}

\author{Kevin B. Stevenson}
\affil{Space Telescope Science Institute, Baltimore, MD 21218, USA}

\author{Andre Wong}
\affil{Jet Propulsion Laboratory, California Institute of Technology, 4800 Oak Grove Drive, Pasadena, California 91109, USA}

\author{John W. Chapman}
\affil{Jet Propulsion Laboratory, California Institute of Technology, 4800 Oak Grove Drive, Pasadena, California 91109, USA}

\author{David R. Ciardi}
\affil{NASA Exoplanet Science Institute, California Institute of Technology, 1200 East California Boulevard, Pasadena, CA 91125, USA}

\author{Andrew W. Howard}
\affil{Department of Astronomy, California Institute of Technology, MC 249-17, 1200 East California Blvd, Pasadena CA 91125, USA}

\author{Tiffany Kataria}
\affil{Jet Propulsion Laboratory, California Institute of Technology, 4800 Oak Grove Drive, Pasadena, California 91109, USA}

\author{Eliza M.-R. Kempton}
\affil{Department of Astronomy, University of Maryland, College Park, MD 20742, USA}

\author{David Latham}
\affil{Harvard-Smithsonian Center for Astrophysics, 60 Garden St., Cambridge, MA 02138}

\author{Suvrath Mahadevan}
\affil{Center for Exoplanets \& Habitable Worlds, The Pennsylvania State University, University Park, PA 16802}
\affil{Department of Astronomy \& Astrophysics, The Pennsylvania State University, 525 Davey Laboratory, University Park, 16802, USA}

\author{Jorge Mel\'{e}ndez}
\affil{Departamento de Astronomia do Instituto de Astronomia, Geof\'{i}sica e Ci\^{e}ncias Atmosf\'{e}ricas, 
Rua do Mat\~{a}o, 1226 - Cidade, Universit\'{a}ria S\~{a}o Paulo-SP - Brasil - 05508-090}

\author{Vivien Parmentier}
\affil{Atmospheric, Oceanic \& Planetary Physics, Department of Physics, University of Oxford, Oxford OX1 3PU, UK}

\vspace{12pt}
\begin{abstract}
Launching in 2028, ESA's 0.64~m$^{2}$ Atmospheric Remote-sensing Exoplanet Large-survey (ARIEL) survey of $\sim$1000 transiting exoplanets will build on the legacies of NASA's Kepler and TESS and complement JWST by placing its high precision exoplanet observations into a large, statistically-significant planetary population context. With continuous 0.5--7.8~$\mu$m coverage from both {{FGS}} ({{0.50--0.6, 0.6--0.81, and 0.81--1.1~$\mu$m photometry; 1.1--1.95~$\mu$m spectroscopy}}) and AIRS (1.95–-7.80~$\mu$m spectroscopy), ARIEL will determine atmospheric compositions and probe planetary formation histories during its 3.5-year mission. NASA’s proposed Contribution to ARIEL Spectroscopy of Exoplanets (CASE) would be a subsystem of ARIEL’s FGS instrument consisting of two visible-to-infrared detectors, associated readout electronics, and thermal control hardware. FGS, to be built by the Polish Academy of Sciences’ Space Research Centre, will provide both fine guiding and visible to near-infrared photometry and spectroscopy, providing powerful diagnostics of atmospheric aerosol contribution and planetary albedo, which play a crucial role in establishing planetary energy balance. The CASE team presents here an independent study of the capabilities of ARIEL to measure exoplanetary metallicities, which probe the conditions of planet formation, and {{FGS}} to measure scattering spectral slopes, which indicate if an exoplanet has atmospheric aerosols (clouds and hazes), and geometric albedos, which help establish planetary climate. Our simulations assume that ARIEL's performance will be 1.3$\times$ the photon noise limit. This value is motivated by current transiting exoplanet observations: Spitzer/IRAC and Hubble/WFC3 have empirically achieved 1.15$\times$ the photon noise limit. One could expect similar performance from ARIEL, JWST, and other proposed future missions such as HabEx, LUVOIR, and Origins. Our design reference mission simulations show that ARIEL could measure the mass-metallicity relationship of its 1000-planet {{single-visit}} sample to $>7.5\sigma$ and that {{FGS}} could distinguish between clear, cloudy, and hazy skies and constrain an exoplanet's atmospheric aerosol composition to $\gtrsim5\sigma$ for hundreds of targets, providing statistically-transformative science for exoplanet atmospheres.

\end{abstract}

\vspace{12pt}
\section{Introduction}
The European Space Agency's (ESA's) Atmospheric Remote-sensing Infrared Exoplanet Large-survey (ARIEL) M4\footnote{the fourth medium-class mission} mission will conduct a visible to infrared (IR) spectroscopic survey with {{its Visible Photometer (VISPhot; visible and near-IR photometry at 0.5--0.6, 0.6--0.81, and 0.81--1.1~$\mu$m), Near-IR Spectrometer (NIRSpec; 1.1–1.95~$\mu$m spectroscopy with a R=20) and the}} ARIEL InfraRed Spectrometer (AIRS: Ch0 from 1.95--3.9~$\mu$m with a R=100 and Ch1 from 3.9--7.8~$\mu$m with a R=30) of known transiting planets \citep{edwards19}. {{For simplicity, we refer to both VISPhot and NIRSpec collectively as the ``FGS instrument'' as they are both part of ARIEL's Fine Guidance System\footnote{Note that the wavelengths used here are adopted from the ARIEL Assessment Study Report, ESA/SCI(2017)2 (http://sci.esa.int/cosmic-vision/59109-ariel-assessment-study-report-yellow-book): VISPhot photometry at 0.50--0.55, 0.6--1.0, and 1.0--1.2~$\mu$m; NIRSpec R=10 spectroscopy from 1.25–1.95~$\mu$m; AIRS spectroscopy  1.95--7.80~$\mu$m with R=30--100}.}} ARIEL, currently scheduled to launch in 2028, will measure their compositions and determine the key factors affecting the formation and evolution of planetary systems \citep{tinetti16, tinetti18}.

\begin{figure*}
\centering
\includegraphics[width=1\textwidth]{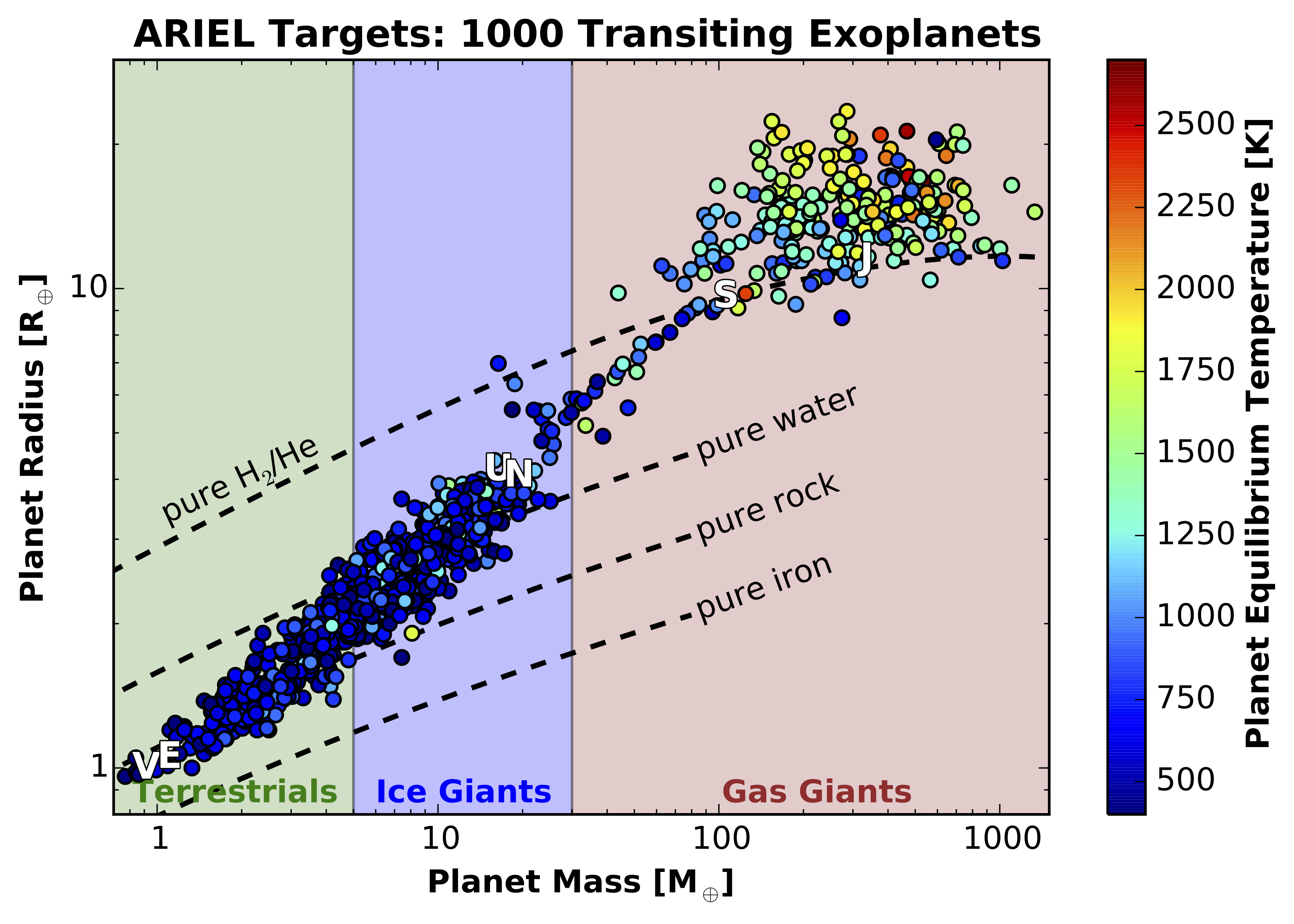}
\caption{{ARIEL will observe a large, diverse target sample}, featuring exoplanets that span a wide range of masses, radii, and equilibrium temperatures. This ARIEL target sample ($\sim$1000 transiting exoplanets) draws from both currently-known exoplanets and those predicted to be discovered by TESS \citep{sullivan15}. Also plotted are mass-radius relationships for planets composed of pure H$_{2}$/He \citep{seager07}, pure water, pure rock, and pure iron \citep{fortney07}.
}
\label{fig:ariel_targets}
\end{figure*}

Over its 3.5 year mission lifetime, the 1.1~m $\times$ 0.7~m ARIEL telescope will observe $\sim$1000 warm and hot transiting planets of diverse {{radii,}} masses, and temperatures (Fig.~\ref{fig:ariel_targets}) using transit, eclipse, and phase curve  observations. Thus, ARIEL will provide a statistically-significant context for the detailed, high-resolution and precision measurements of tens of transiting exoplanets by the James Webb Space Telescope \citep[JWST;][]{cowan15}. The ARIEL data will map the relationship between atmospheric metallicity and planet mass, which is essential to understanding planet formation, and probe atmospheric loss in the critical sub-Neptune to terrestrial region \citep[e.g.,][]{owenwu13, fulton17, swain18}. Our Solar System’s planets barely hint at the diversity that is just emerging from exoplanet observations, which indicate, for example, widely varying metal-enrichments at a given mass \citep{kreidberg14, morley17,wakeford17b}. However, the few atmospheric metallicity measurements largely derive from water vapor, due to Hubble/WFC3’s wavelength limitations, unless additional observation modes or platforms provide wavelength coverage \citep[e.g., with Hubble/STIS or Spitzer/IRAC;][]{sing16, stevenson17, wakeford17, nikolov18, wakeford18, alam18, pinhas19}. For exoplanets with limited wavelength coverage (e.g., Hubble/WFC3 alone), a solar abundance ratio for O and C is typically assumed because the measurements are not sensitive to all the chemical species that carry the major elements \citep[e.g.,][]{kreidberg14b}. An inventory of the key molecules that form from the cosmically most abundant elements ({{H, He, O and C, which at the temperatures and pressures of interest in, e.g., disks and exo-atmospheres, lead to the formation of}} namely H$_{2}$O, CH$_{4}$, CO, and CO$_{2}$), particularly in the infrared where these species strongly absorb \citep{seager00}, is needed to quantify atmospheric composition and understand the diversity of exoplanets.


\begin{figure}[!htb]
\centering
\includegraphics[width=1\columnwidth]{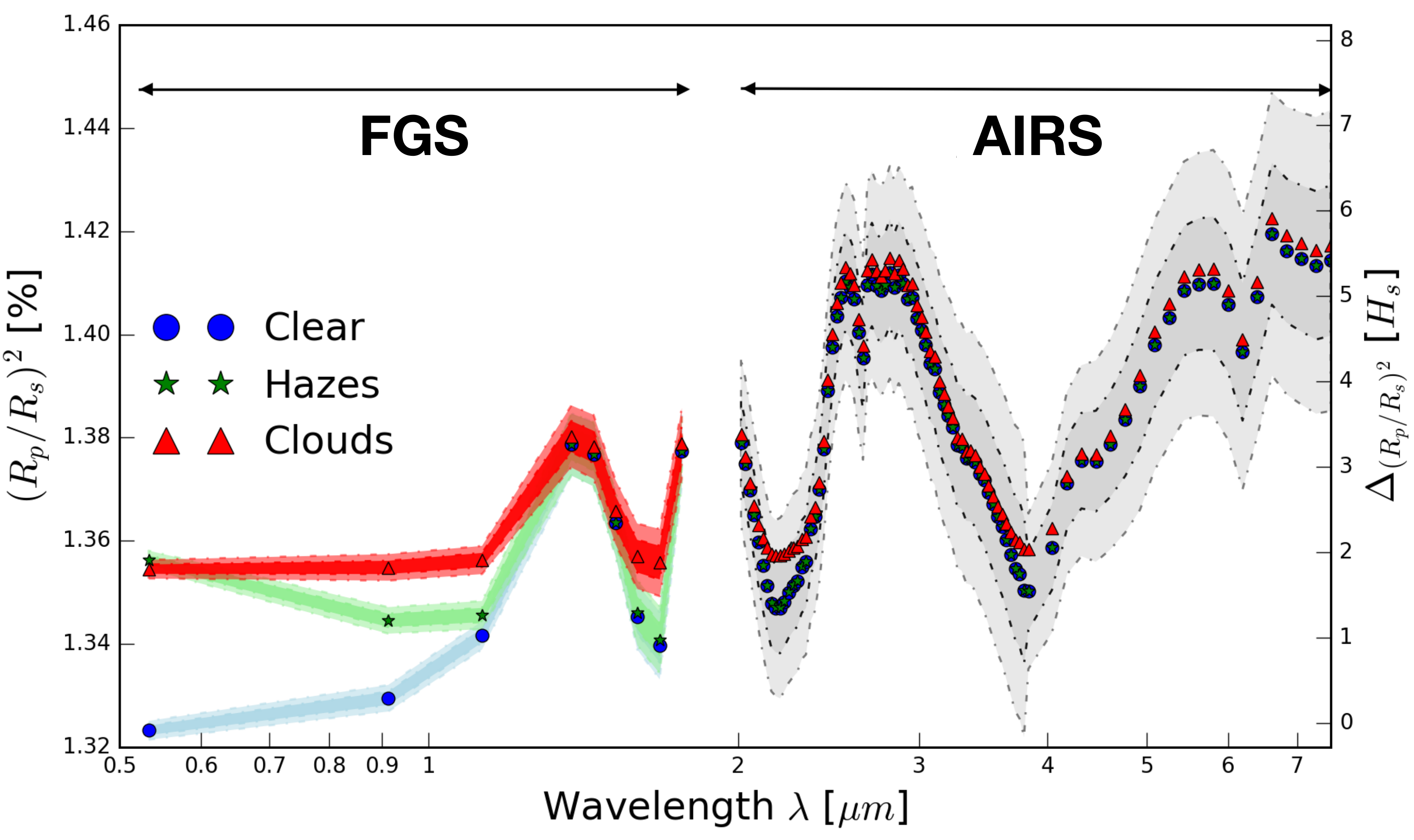}
\caption{{{{FGS}} will distinguish between clear, cloudy, and hazy worlds} by measuring visible to near-IR scattering spectral slopes during transit. In doing so, {{FGS}} will reduce degeneracies between molecular mixing ratios and cloud heights and characteristics as measured by AIRS in the infrared. {{Shown are simulations for a Type 3 (10 visits) WASP-12b-like planet whose atmosphere is dominated by H$_{2}$, He, and H$_{2}$O. The blue curve assumes a semi-finite bottom cloud, completely opaque, with a cloud top at 10 bars and 100~ppm of H$_{2}$O. The green curve assumes a uniformly distributed haze layer on the top of the bottom cloud \citep[scaling of H$_{2}$ Rayleigh scattering;][]{etangs08} on the top of the blue model parameters. The red curve assumes semi-finite bottom cloud, completely opaque, with a cloud top at 10~mbars and 100~ppm of H$_{2}$O.}}
}
\label{fig:xo2}
\end{figure}

{{ARIEL/FGS}} will determine exoplanet aerosol properties by measuring the light scattered by the exoplanetary atmospheres during transit and geometric albedos during eclipse. Haze and clouds play a fundamental role in the energy balance of exoplanets. In addition they mask molecular spectral features measured by AIRS in the infrared (Fig.~\ref{fig:xo2}). Generally, clouds, which form from condensation of atmospheric constituents, dampen spectral modulation in both the visible and IR (across the {{FGS}} and AIRS bandpasses). Hazes, which form from photochemistry, typically dampen features in the visible and near-IR ({{FGS}}) only \citep{tinetti10}. Thus, cloudy exoplanets will generally have similar transit depths in the near- and mid-IR ({{FGS}} and AIRS ranges) while hazy objects will have larger transit depths in the visible and near-IR ({{FGS}}) than the mid-IR (AIRS). Thus {{FGS}} could determine if an atmosphere is clear, cloudy, or hazy (Fig.~\ref{fig:xo2}) and retrieve the location of its optically-thick radius. Such observations are crucial as even a $\sim$1\% difference in the estimated planetary radius at the $\sim$1 bar pressure level can result in variations in molecular abundances of an order of magnitude or more \citep{tinetti10}. Thus, the combination of {{FGS}} and AIRS data provide ARIEL a robust estimate of physical quantities such as molecular mixing ratios or cloud top pressure \citep{tinetti10, benneke12, benneke13, griffith14, betremieux14, betremieux15, zellem15, betremieux16, betremieux17}.

The NASA Contribution to ARIEL Spectroscopy of Exoplanets (CASE), {{an Explorer-class Mission of Opportunity currently in Step~2 Review,}} would provide {{the two visible-to-infrared Teledyne Imaging Sensors Sensor Chip Assemblies \citep[SCAs;][]{loose03, beletic08}, associated readout electronics, and thermal control hardware as a subsystem of ARIEL's FGS instrument, which will be built by the Space Research Centre from the Polish Academy of Sciences. The SCAs serve a dual purpose as fine guidance sensors}} and provide the short wavelength coverage for studies of exoplanet clouds and hazes. Both AIRS and {{FGS}} simultaneously observe each ARIEL target; thus the combination of {{FGS}} and AIRS provides ARIEL simultaneous 0.5--8~$\mu$m coverage. The broader {{FGS}} science themes do not drive ARIEL operations. Rather, ARIEL places requirements fine guiding that are flowed down to establishing {{FGS's}} science capabilities and margins. {{The ARIEL fine guiding functions places three key requirements on the CASE contribution: the bandpass, line of sight measurement noise (astrometric precision), and frame rates. Of the two FGS bandpasses, 0.8--1.0~$\mu$m and 1.0--1.2~$\mu$m, only one channel is needed and the second provides redundancy. For bright stars, the line of sight noise shall be $\le$20~mas and needs to support 10~Hz frame rates for both cases.\footnote{M4 Mission Selection Review Board Report (public) (https://www.cosmos.esa.int/documents/1365222/1365271/M4+MSR+Board+Report+ESA-SCI-F-ESTEC-RP-2017-006+Public+Issue+1-0+Final.pdf/401667c7-a638-c2ba-3cae-85c154e4b7e9)}}}

Here we describe {{an independent study by the CASE team of ARIEL's performances, specifically its ability to measure exoplanetary metallicities and FGS's}} ability to measure geometric albedos and aerosol scattering spectral slopes.

\section{Performance Simulations}
\subsection{Simulating the ARIEL Target Population}
The ARIEL survey is organized into a three-type approach (ARIEL Assessment Study Report, ESA/SCI(2017)2\footnote{\label{yellowbook}http://sci.esa.int/cosmic-vision/59109-ariel-assessment-study-report-yellow-book/}). Type~1 is a $\sim$1000 planet ``Reconnaissance survey'' to conduct a population analysis to determine, for example, what fraction of the sample has atmospheric aerosols (clouds and hazes) that obscure IR molecular spectral modulation; such targets will be excluded from subsequent Types. For Type~1, the CASE Team independently assumed 1 transit per target; this assumption does not necessarily follow the current ARIEL 3-Type strategy. Type~2 is a $\sim$500 planet ``Deep survey'' featuring multiple ($\sim$5) revisits to measure molecular abundances of relatively clear Type~1 targets. Type 3 is a detailed study of $\sim$50 ``Benchmark planets'' featuring more visits ($\sim$10) to conduct in-depth analyses of high-priority targets, as deemed by the ARIEL Consortium. The number and breakdown of transmission, emission, and phase curve targets will be determined by the scientific priorities established by the ARIEL Consortium, with input from the CASE science team. For comparison, a recent study by the ARIEL Consortium presents a three-tiered observing strategy where Tier 1 includes 1000 planets, Tier 2 includes 600 planets, and Tier 3 includes 50 planets; the number of revisits in each Tier is motivated by the observed precision on a target-by-target basis \citep{edwards19}.


We simulate the ARIEL target sample by drawing from both the currently-known transiting exoplanets \citep[via the NASA Exoplanet Archive;][]{akeson13} and those predicted to be discovered with TESS \citep{sullivan15}. To cull these two lists of thousands of planets down to the 1000 planet ARIEL sample, we apply two platform-independent figures of merit (FOM) that provide a relative ranking of both transit and eclipse targets. To determine ARIEL's targets for transit observations, we use the FOM as defined in \citet{zellem17}:
\begin{align}
\mathrm{FOM_{transit}} &= \frac{\mathrm{signal}}{\mathrm{noise}} \nonumber \\ 
&= \frac{\mathrm{1 \ H_{s} \ of \ spectral \ modulation}}{\mathrm{photon \ noise}} \nonumber \\ 
&= \frac{2H_{s}R_{p}R_{s}^{-2}}{10^{0.2 H\text{-}mag}}
\label{eqn:transit_FOM}
\end{align}
where $R_{p}$ is the planet's radius, $R_{s}$ is the host star's radius, $H$-$mag$ is the host star's apparent magnitude in the H-band, and $H_{s}$ is an exoplanet's atmospheric scale height:
\begin{align}
H_{s} = \frac{k_{B}T_{eq}}{\mu g}
\end{align}
where $k_{B}$ is the Boltzmann constant, $T_{eq}$ is the planet's equilibrium temperature, $\mu$ is the mean molecular mass of the planet's atmosphere, and $g$ is the planet's acceleration due to gravity. (Note that this FOM can also be fine-tuned to better match any proposed transit observations; for example, if one were to observe in the K-band, then they would use the star's $K$-$mag$ in place of $H$-$mag$ in Equation~\ref{eqn:transit_FOM}.) Similar figures of merit have been independently derived by \citet{cowan15}, \citet{goyal18}, \citet{kempton18}, and \citet{morgan18}.

To determine the best targets for eclipse and phase curve observations, we produce an additional ranking of the 1000 transit targets with the FOM as defined in \citet{zellem18}:
\begin{align}
\mathrm{FOM_{eclipse}} &= \frac{\mathrm{signal}}{\mathrm{noise}} \nonumber \\
&= \frac{\mathrm{eclipse \ depth}}{\mathrm{photon \ noise}} \nonumber \\
&= \frac{F_{p}R_{p}^{2}F_{s}^{-1}R_{s}^{-2}}{10^{0.2 H\text{-}mag}}
\label{eqn:eclipse_FOM}
\end{align}
where $F_{p}$ and $F_{s}$ are the fluxes of the planet and host star in the H-band, respectively. For simplicity's sake, we assume that to first order both the star and planet radiate as perfect blackbodies and thus calculate $F_{p}$ and $F_{s}$ via the Planck function and the equilibrium and effective temperatures of the planet and star, respectively. (Please note that this FOM can also be fine-tuned to better match the proposed observations by choosing a different photometric bandpass for the host star's magnitude and by calculating the planet-to-star flux ratio at a different wavelength.) A similar figure of merit has been independently devised by \citet{cowan15} and \citet{kempton18}.

All of the input parameters for these two FOMs were taken from NASA's Exoplanet Archive \citep{akeson13} and \citet{sullivan15}, except the planet's mean molecular mass $\mu$. {{As explained in detail in \citet{zellem17}}}, for this parameter, we calculate a mean molecular weight $\mu$ for each planet using an assumed metallicity-mass relationship {{motivated both by simulations of planet formation \citep{fortney13} and by observations of methane in the Solar System's giant planets \citep{wong04, fletcher09, karkoschka11, sromovsky11} and a C/O ratio drawn from a Gaussian distribution. Using this metallicity and C/O ratio, we calculate the full range of molecular abundances within each planet's atmosphere assuming equilibrium chemistry \citep{mcbride96} and translate these abundances into an overall mean molecular weight $\mu$ for each atmosphere.}}

Other than defining an observing band, both of these figures of merit are independent of the observing platform, allowing one to quickly rank all known transiting exoplanets for any telescope. These FOMs will particularly be useful for culling the thousands of exoplanets discovered with campaigns like TESS \citep{ricker14} for follow-up with both ground- and space-based observatories, such as the James Webb Space Telescope (JWST), ARIEL, or any of the mission concepts currently under Astro2020 decadal review that could conduct transit observations, such as HabEx \citep{gaudi18}, LUVOIR \citep{luvoir18}, and the Origins Space Telescope \citep{battersby18}.

\subsubsection{Stellar Activity}
{{Stellar activity can affect a transiting exoplanet’s observed spectrophotometric signal and potentially alter its inferred physical properties \citep[e.g,][]{pont08, silvavalio08, czesla09, wolter09, agol10, berta11, carter11, desert11, sing11, fraine14, mccullough14, oshagh14, barstow15, damasso15, zellem15, zellem17, rackham17, rackham18, morris18, pinhas18, espinoza19}. However, it has been found that stellar variability has a negligible effect on epoch-to-epoch transit observations, particularly in the infrared \citep[e.g.,][]{fraine13, kreidberg14, zellem17, morris18, bruno18, kilpatrick19}, such as those that will be conducted by ARIEL \citep{zellem17}. Therefore, based on these studies, we assume that stellar variability has a negligible effect on our simulated ARIEL observations.

{{However, stellar activity could affect the FGS visible and near-IR wavelengths, especially for bright, very active stars, thereby falsely altering the physical interpretation about the planet's atmosphere, such as the existence of hazes \citep[e.g.,][]{mccullough14}. While we intend to explore the assumption that stellar activity is negligible across all ARIEL wavelengths for the vast majority of its targets in a future, in-depth study, we note that there are multiple efforts to treat stellar contamination that could be used to correct the ARIEL/FGS data \citep{rackham17, rackham18, pinhas18, espinoza19}.}}

\subsection{Calculating ARIEL's Signal-to-Noise Ratios}
We adapt an existing radiometric model \citep[described in detail in][{{this radiometric model has been validated against existent Hubble/WFC3 observations and is capable of including noise sources such as shot noise, the zodiacal background, background noise from the telescope, background noise from the instrument, detector dark current, and read-out noise}}]{chapman17} to match ARIEL's required performance. {{The CASE Team's}} radiometric model {{ has been qualitatively verified with the ARIEL Consortium's precision estimates \citep{tinetti18}.}} {{(The performances used here reflect those illustrated in the ARIEL Assessment Study Report, ESA/SCI(2017)2\textsuperscript{\ref{yellowbook}} and not necessarily the latest performance models by the ARIEL Consortium.)}} As indicated by its bright- and dim-stressing cases, all ARIEL observations are photon noise dominated (see Fig.~4-28 in the ARIEL Assessment Study Report, ESA/SCI(2017)2\textsuperscript{\ref{yellowbook}}).

We use the {{CASE Team's}} radiometric model to calculate the signal-to-noise ratios (SNRs) for {{FGS}} and AIRS for every ARIEL target. We further degrade these required SNRs by 30\%; this margin is motivated by current transiting exoplanet post processing capabilities for both Hubble/WFC3 spatial scanned and Spitzer/IRAC full-orbit phase curve observations, which achieve on average 1.3$\times$ and 1.14$\times$ the photon noise limit, respectively (Tables~$\ref{table:wfc3_targets}$ and $\ref{table:spitzer_targets}$; Fig.~\ref{fig:SNR_observed_vs_ideal}). These values are consistent with previous studies: while \citet{hansen14} suggested that the first generation of Spitzer eclipse studies had underestimated uncertainties, \citet{ingalls16} demonstrated that repeat observations reduced and analyzed with the latest techniques yield eclipse uncertainties near the photon noise limit. In addition, the Spitzer photon noise factor we report here is consistent with a recent study of 78 Spitzer eclipses \citep{garhart19}. When removing the outlier of WASP-103b \citep[3.7$\times$ the photon noise;][]{cartier17}, Hubble/WFC3 achieves 1.13$\times$ the photon noise limit, bringing it more in-line with Spitzer/IRAC's performance. These Hubble/WFC3 and Spitzer/IRAC data indicate that transit observations can currently achieve an average precision down to 15~ppm \citep[with Hubble/WFC3;][]{tsiaras16, line16} and have not yet reached a noise floor (Fig.~\ref{fig:SNR_observed_vs_ideal}).

These Hubble and Spitzer data indicate that research groups are capable of modeling detector trends \citep[some studies have been able to get within 1.05$\times$ WFC3's photon noise limit; e.g.,][]{tsiaras16, kreidberg18} as new sources of detector systematic errors \citep{rauscher07} are revealed by ever-higher dynamic range observations \citep[e.g., using the pixel-mapping technique to correct for warm Spitzer's intrapixel effect;][]{ballard10}. The combination of repeated observations and strong astrophysical priors allows groups to {{successfully model systematic}} errors and achieve near photon-limited performance, as is empirically shown by the publication track record \citep[e.g.,][]{cowan12, knutson12, deming13, lewis13, maxted13, wakeford13, crouzet14, fraine14, knutson14a, knutson14b, kreidberg14, zellem14, haynes15, sing15, wong15, demory16, evans16, sing16, tsiaras16, tsiaras16b, wong16, cartier17, damiano17, evans17, stevenson17, wakeford17, alam18, bruno18, dang18, kilpatrick18, nikolov18, wakeford18, zhang18}.

We assume that this trend will continue, suggesting that future observations with JWST and ARIEL, and proposed NASA missions, like HabEx, and LUVOIR, and the Origins Space Telescope, when coupled with post processing and multiple revisits (as is the current plan for ARIEL's Type 2 and 3 planets), could achieve observations near the photon noise limit and $<$15~ppm.

{{Applying the analysis that was used to predict ARIEL’s performance can also be used to predict the performance for JWST, which will be making complementary, in depth observations complementary to ARIEL.}} Thus the presumed noise floors of 20, 30, and 50~ppm adopted for JWST's NIRISS SOSS, NIRCam grism, and MIRI LRS by \citet{greene16} could prove to be conservative. Given that NIRCam, NIRISS, and NIRSpec also use a HgCdTe detector like WFC3 \citep{beichman12, doyon12, teplate05} and JWST's larger aperture size, we thus scale the currently-best WFC3 precision \citep[15~ppm;][]{tsiaras16} by the ratio of JWST's to Hubble's primary mirror sizes to estimate that the noise floor for NIRISS SOSS could be as low as 5.3~ppm (assuming photon-noise limited observations), 6.0~ppm (assuming 1.14$\times$ the photon noise), or 7.0~ppm (assuming 1.33$\times$ the photon noise). Assuming similar reductions in the noise floors of JWST's other instruments, we predict that NIRCam grism could have a noise floor of 8.0~ppm (assuming photon-noise limited observations), 9.1~ppm (assuming 1.14$\times$ the photon noise), or 10.6~ppm (assuming 1.33$\times$ the photon noise) and MIRI LRS could have a noise floor of 13.3~ppm (assuming photon-noise limited observations), 15.2~ppm (assuming 1.14$\times$ the photon noise), or 17.6~ppm (assuming 1.33$\times$ the photon noise). We note that since these values are calculated by extrapolating from the current highest precision with WFC3, they could prove to be over-estimated for JWST.


\begin{table*}[htb!]
\resizebox{1\textwidth}{!}{%
\begin{tabular}{llllllll}
Planet              & Obs. & Peak-to-Valley  & Average & Average Per-visit  & Modeling & Photon                 & Reference              \\

              &  Type & of Extracted  & Spectral & Spectral & Gain & Noise                &               \\

              &  & Light Curve {[}ppm{]} & Uncertainty {[}ppm{]} & Uncertainty {[}ppm{]} &  & Factor {[}\%{]}     &               \\
\hline
\hline
55 Cnc e            & Transit   & 2000                                              & 15                                     & 21                                               & 95            & 105                          & \citet{tsiaras16}    \\
GJ 1214b            & Transit   & 4274                                              & 28                                     & 97                                               & 44            & 114                          & \citet{kreidberg14}  \\
GJ 436b             & Transit   & 2000                                              & 43                                     & 86                                               & 23            & 120                          & \citet{knutson14a}    \\
HAT-P-1b            & Transit   & 4000                                              & 161                                    & 161                                              & 25            & 112                          & \citet{wakeford13}   \\
HAT-P-11b           & Transit   & 1000                                              & 45                                     & 45                                               & 22            & Not Reported                 & \citet{fraine14}     \\
HAT-P-32b           & Transit   & 2500                                              & 114                                    & 114                                              & 22            & 110                          & \citet{damiano17}    \\
HAT-P-38b           & Transit   & 2000                                              & 89                                     & 126                                              & 16            & Not Reported                 & \citet{bruno18}      \\
HD 97658b           & Transit   & 1000                                              & 24                                     & 24                                               & 42            & 105                          &\citet{knutson14b}   \\
HD 189733b          & Eclipse   & 1721                                              & 58                                     & 58                                               & 30            & Not Reported                 & \citet{crouzet14}    \\
HD 209458b          & Transit   & 2000                                              & 36                                     & 36                                               & 56            & 126                          & \citet{deming13}     \\
HD 209458b          & Transit   & 2424                                              & 41                                     & 41                                               & 59            & Not Reported                 & \citet{tsiaras16b}   \\
WASP-31b            & Transit   & 2000                                              & 271                                    & 271                                              & 7             & 147                          & \citet{sing15}       \\
WASP-33b            & Eclipse   & 2000                                              & 25                                     & 35                                               & 57            & 105                          & \citet{haynes15}     \\
WASP-39b            & Transit   & 1000                                              & 180                                    & 255                                              & 4             & Not Reported                 & \citet{wakeford18}   \\
WASP-63b            & Transit   & 3250                                              & 49                                     & 49                                               & 66            & Not Reported                 & \citet{kilpatrick18} \\
WASP-67b            & Transit   & 2500                                              & 243                                    & 243                                              & 10            & Not Reported                 & \citet{bruno18}      \\
WASP-101b           & Transit   & 2000                                              & 70                                     & 70                                               & 29            & Not Reported                 & \citet{wakeford17}  \\
WASP-103b           & Eclipse   & 3000                                              & 175                                    & 247                                              & 12            & 370                          & \citet{cartier17}    \\
WASP-121b & Transit   & 1667                                              & 100                                    & 100                                              & 17            & 105                          & \citet{evans16}      \\
WASP-121b           & Eclipse   & 2000                                              & 89                                     & 89                                               & 22            & 104                          & \citet{evans17}      \\
XO-1b               & Transit   & 3000                                              & 96                                     & 96                                               & 31            & 106                          & \citet{deming13}     \\
\hline
AVERAGE             &           & 2254                                              & 93                                     & 108                                              & 33            & 133                          &                                      
\end{tabular}%
}
\caption{\textbf{Published Hubble/WFC3 Scan-mode Transit Spectroscopy To Date.} On average, Hubble/WFC3 post processing can reduce systematic errors by a factor of 33, getting within 33\% of the photon noise limit. When removing WASP-103b, Hubble/WFC3 can achieve 1.13$\times$ the photon noise limit, which brings it more in-family with Spitzer/IRAC (Table~\ref{table:spitzer_targets}). Please note that we omit any analyses that do not include their raw lightcuves as these data are necessary to calculate the ``Modeling Gain'': the reduction of noise sources via the ratio of the peak-to-valley amplitude of the extracted raw light curve to the average per-visit uncertainty on the final, reduced spectrum.}
\label{table:wfc3_targets}
\end{table*}

\begin{table*}[htb!]
\resizebox{\textwidth}{!}{%
\begin{tabular}{lllllll}
Planet     & Peak-to-Valley & Phase Curve  & Modeling & Photon                & Reference           \\
     & of Extracted  & Amplitude & Gain & Noise                &            \\
     & Light Curve {[}ppm{]} & {[}ppm{]} &  & Factor {[}\%{]} &                &           \\
\hline
\hline
55 Cnc e   & 10000                                             & 34                    & 294           & Not Reported                 & \citet{demory16}  \\
CoRoT-2b   & 35000                                             & 200                   & 175           & Not Reported                 & \citet{dang18}    \\
HAT-P-2b (3.6~$\mu$m)  & 20000                                            & 1140                    & 18           & 105                          & \citet{lewis13}    \\
HAT-P-2b  (4.5~$\mu$m) & 15000                                             & 790                    & 19           & 111                          & \citet{lewis13}    \\
HAT-P-7b   & 60000                                             & 95                    & 632           & 112                          & \citet{wong16}    \\
HD 149026b & 40000                                             & 189                   & 212           & 128                          & \citet{zhang18}   \\
HD 189733b & 20000                                             & 61                    & 328           & 112                          & \citet{knutson12} \\
HD 209458b & 25000                                             & 115                   & 217           & 114                          & \citet{zellem14}   \\
WASP-12b   & 14500                                             & 300                   & 48            & Not Reported                 & \citet{cowan12}   \\
WASP-14b   & 40000                                             & 39                    & 1026          & 114                          & \citet{wong15}    \\
WASP-18b   & 20000                                             & 219                   & 91            & Not Reported                 & \citet{maxted13}  \\
WASP-19b   & 20000                                             & 175                   & 114           & 115                          & \citet{wong16}    \\
WASP-33b   & 60000                                             & 936                   & 64            & 112                          & \citet{zhang18}   \\
\hline
AVERAGE    & 29192                                             & 330                   & 249           & 114                          &                                    
\end{tabular}%
}
\caption{\textbf{Published Spitzer/IRAC Full-orbit Phase Curves To Date.} On average, Spitzer/IRAC post processing can reduce systematic errors by a factor of 249, getting within 14\% of the photon noise limit. Here we define the ``Modeling Gain'' as the reduction of noise sources via the ratio of the peak-to-valley amplitude of the extracted raw light curve to the uncertainty on the final, reduced phase curve amplitude.}
\label{table:spitzer_targets}
\end{table*}

\begin{figure}[!htb]
\centering
\includegraphics[width=1\columnwidth]{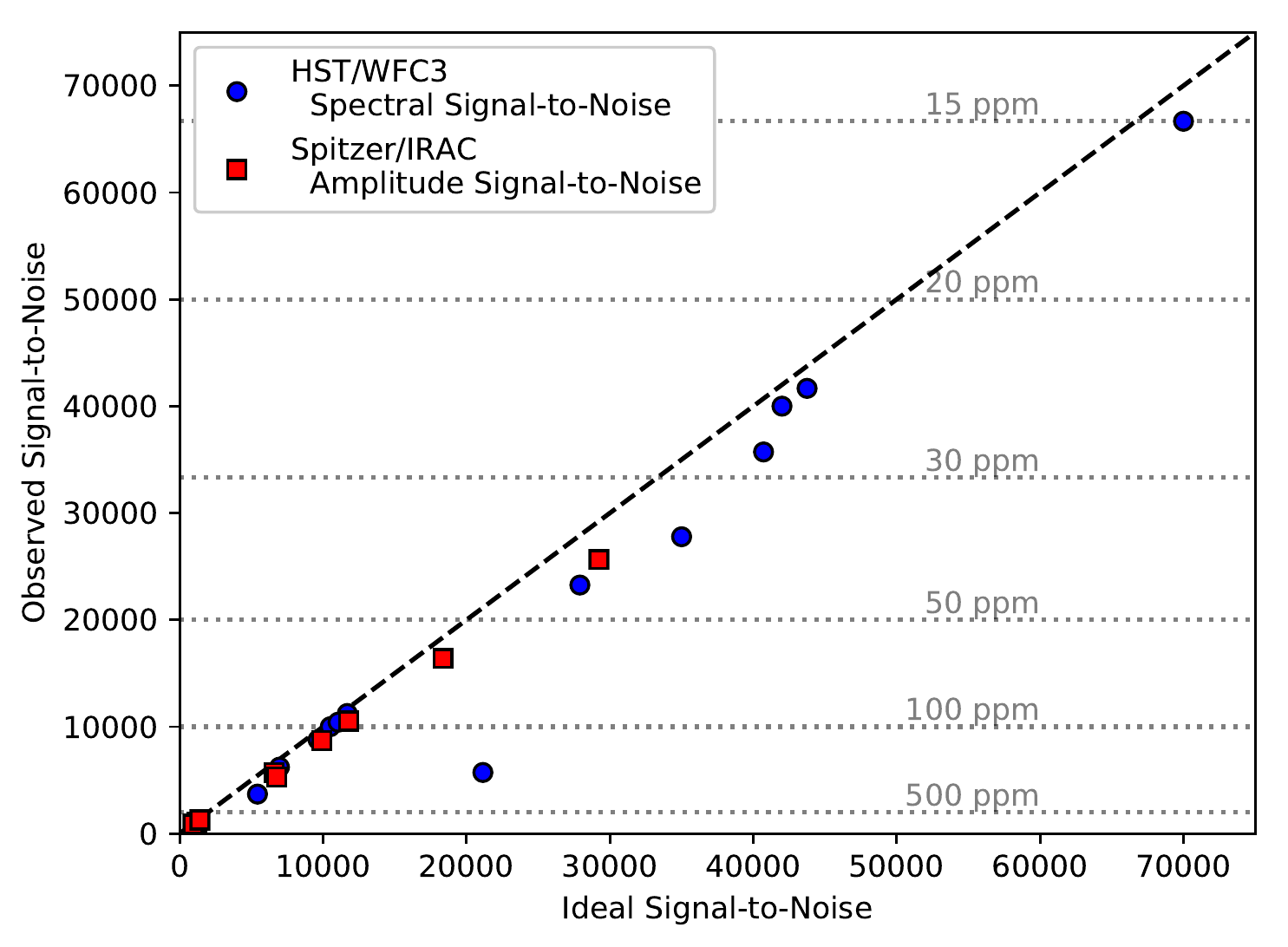}
\caption{{Hubble/WFC3 and Spitzer/IRAC transiting exoplanet observations have not yet reached a noise floor.} A noise floor would manifest as a turnover point or ceiling whereby the maximum observed SNR would remain constant despite increasing ideal (Poisson) SNR. (Data from Tables~\ref{table:wfc3_targets} and \ref{table:spitzer_targets}.)
}
\label{fig:SNR_observed_vs_ideal}
\end{figure}

%
%
%
%
%

\subsection{Metallicity Simulations}
A general outcome of giant planet formation models is the relation between planet mass and atmospheric metallicity, with lower mass planets having higher metallicities. \citet{fortney13} find a nearly linear relationship in log(mass) for higher-mass planets, along with a change in slope at lower masses (Fig.~\ref{fig:mass_met_4panel}).  Furthermore, there is a spread in metallicity at every mass, and their simulations found two different mass-metallicity relations for two different assumed planetesimals sizes. Thus, ARIEL measurements of the mass-metallicity relation will directly impact our understanding of the amount of solids accreted, the efficiency of the vaporization of solids in the atmosphere, and the size of the accreted planetesimals. {{There are a wide range of predictions for metal-enrichment and atmospheric abundance ratios, from various flavors of core-accretion planet formation \citep[e.g.,][]{mordasini16, espinoza17, madhu17}, which need observational constraints.}}

\begin{figure*}[!htb]
\centering
\includegraphics[width=.8\textwidth]{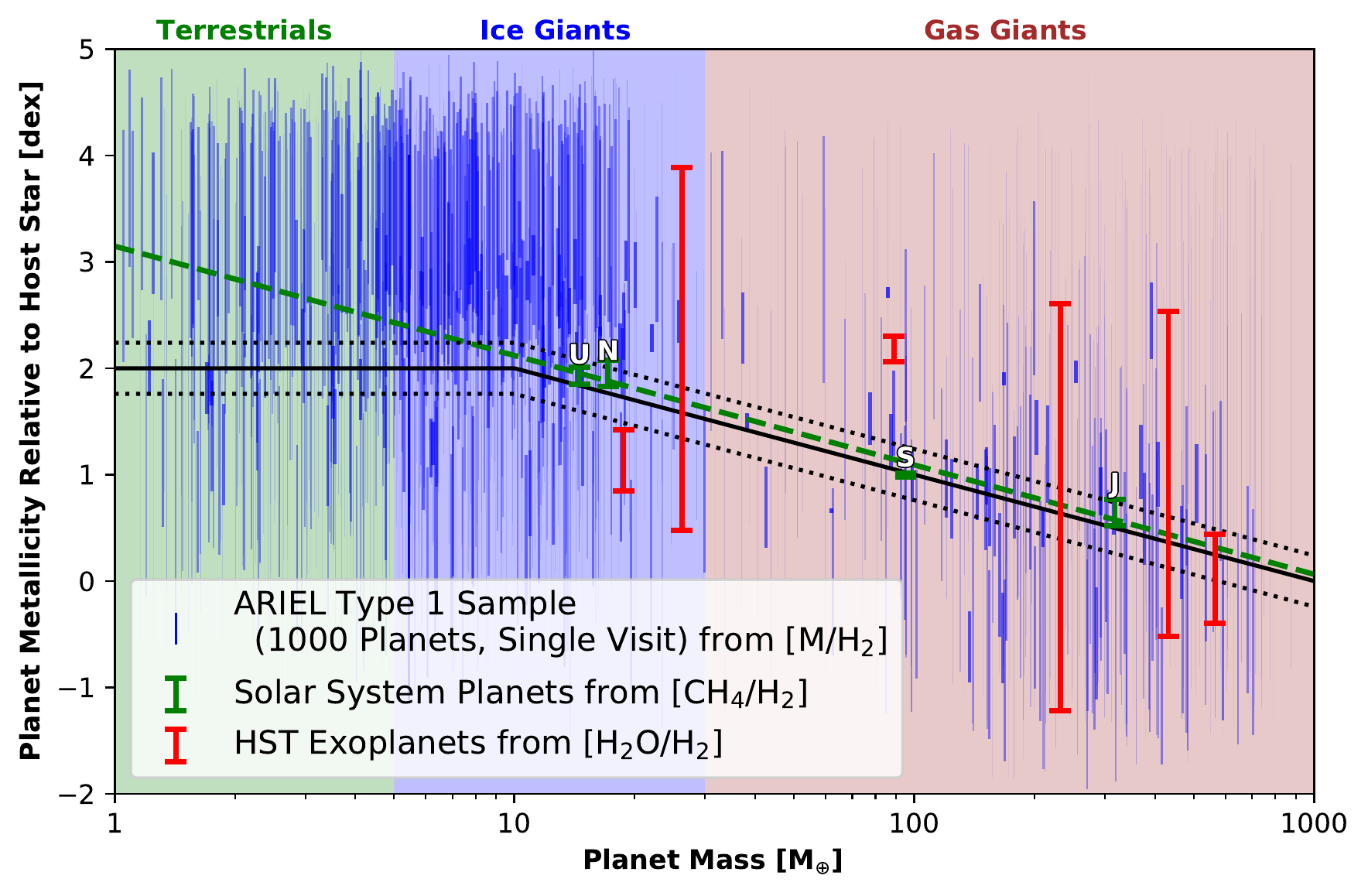}
\includegraphics[width=.8\textwidth]{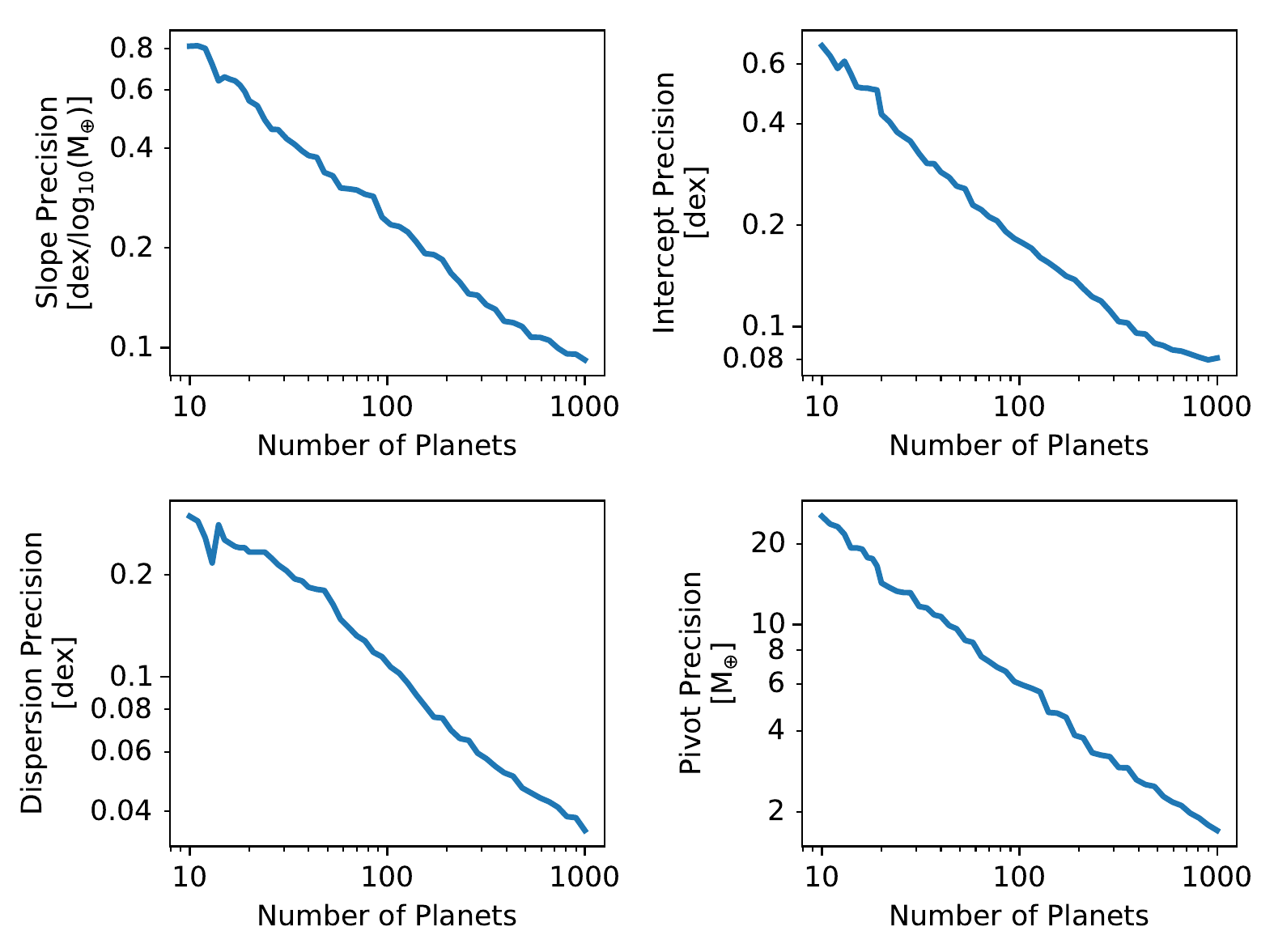}
\caption{{ARIEL will characterize the mass-metallicity relation (top panel) and determine if the trend observed in the Solar System is a universal outcome of the planet formation process.} The Solar System giants (green with a dashed power law fit) and the six measured exoplanets (red) \citep{fraine14, kreidberg14b, kreidberg15, line16, wakeford17, wakeford17b} are shown. {{ARIEL's Type 1 Sample (1000 planets, 1 visit each; simulated in blue)}} will determine the key diagnostic parameters of the mass-metallicity relation identified by theoretical models: offset, slope, spread, and breakpoint (black lines: model fit with $\pm1\sigma$ uncertainty to the simulated data; retrieved values in four panels below).
}
\label{fig:mass_met_4panel}
\end{figure*}

By performing a large, uniform transmission survey with {{FGS}} and AIRS, ARIEL will reveal the slope, vertical offset (the intercept), and distribution (spread) of the mass-metallicity relation as a function of planet mass, and a hypothesized break (pivot) in slope near the low-mass end \citep[Fig.~\ref{fig:mass_met_4panel};][]{fortney13, kreidberg14, thorngren18}. For these metallicity simulations, we use the chemically consistent version of CHIMERA \citep{line14, line15, swain14, kreidberg15} in which molecular abundances are calculated assuming local thermal equilibrium \citep{fegley94, lodders02} and radiative transfer is performed assuming spherical symmetry (1-D). We pre-calculated a library of CHIMERA spectra and then interpolate within this grid. This interpolation scheme allows us to efficiently perform trade studies, where we quantify how science performance varies as a function of mission requirements.

{{We use the same chemically consistent CHIMERA retrieval forward model as the forward model used to generate the synthetic data. In that sense, the constraints are representative within the context of that specific model. This chemically consistent CHIMERA model also includes an approximation to vertical mixing via a quench pressure approximation (2 free parameters) for the Nitrogen and Carbon species \citep[as described in][]{kreidberg18}. Within our framework, vertical mixing is crudely ``marginalized''. While we do not know the degree to which disequilibrium chemistry will manifest itself, we imagine in a ``simplistic'' sense that accounting for, say photochemistry, would require the inclusion of additional free parameters. More free parameters, in general, lead to more degeneracy ``inflating'' uncertainties on the already included parameters. This effect is not likely to be orders of magnitude but should be rather small.  For example, it has been repeatedly demonstrated that photochemistry does little to impact the major O-C-N reservoirs \citep[e.g.,][and references thereafter]{line11, moses13}. The likely consequence is the ``presence'' of trace species like HCN, C$_{2}$H$_{2}$, C$_{2}$H$_{6}$, etc., which in principle could bias a retrieval if such species are present and not accounted for.  Such a bias could be degenerate with, say, high C/O. However, the presence of H$_{2}$O and these species would be more of an indication of photochemistry rather than high C/O, thus, such a case would indicate the need for photochemistry in the retrieval parameterization. Vertical mixing even, in the ``cool planet regime'' ($\lesssim$800K), also does very little to perturb the major C-N-O reservoirs as they are already rather abundant and have nearly constant with altitude vertical mixing ratio profiles anyway.}}

For each of the 1000 Type~1 simulated ARIEL transmission spectra (assuming 1 visit per target), CHIMERA extracts several key planetary parameters---optically-thick radius, temperature profile, metallicity, gravity, and C/O ratio---by matching each simulated planetary spectrum with its best fit in the spectral library. The spectral retrieval also simultaneously fits for the clouds and haze that can blur the information content of each planet's atmosphere. Our prescription for clouds is motivated by a HST/WFC3 study of 19 transiting planets finding a wide range of cloud obscuration for the water feature at 1.3~$\mu$m \citep{iyer16}.  We similarly choose a range of cloud-top pressures (log(P$_{c}) = -2.5 \pm 0.5$~bars) such that some atmospheres are relatively clear and others are more obscured as observed during transit.



\subsection{Aerosol Scattering Slope Simulations}
We define {{FGS's}} capability to measure atmospheric aerosols by establishing its ability to measure a planet's visible-to-near-IR absorption spectral slope across its FGS1 {{(the 0.52 and 0.90~$\mu$m photometric bands) and FGS2 (the 1.12~$\mu$m photmetric band and the 1.25--1.95~$\mu$m spectrograph)}}. A planet's spectral slope observed during transit crucially allows one to distinguish between a clear, cloudy, and hazy atmosphere. For example, a flat spectrum could indicate clouds masking molecular features, as is the case for the cloudy-archetype GJ~1214b \citep{kreidberg14}. A planet with a clear atmosphere will feature a spectral slope at visible wavelengths that follows a Rayleigh trend while a planet with hazes will have a comparatively steeper spectral slope, as is the case for HD~189733b \citep{pont08, pont13, sing16}.


To determine {{FGS's}} ability to measure and distinguish between a clear, cloudy, and hazy atmosphere, we calculate its uncertainty on measuring a planet's visible-to-near-IR spectral slope $\Delta\Big(\frac{R_{p}^{2}}{R_{s}^{2}}\Big)/\Delta\lambda$. For each planet in ARIEL's 1000~target Type 1 sample, we calculate the uncertainty of measuring a planet's transit depth averaged across each of {{FGS's}} FGS1 (the 0.52 and 0.90~$\mu$m photometric bands) and FGS2 (the 1.12~$\mu$m photmetric band and the 1.25--1.95~$\mu$m spectrograph). These two band-averaged uncertainties are then combined to estimate $\Delta\Big(\frac{R_{p}^{2}}{R_{s}^{2}}\Big)$; $\Delta\lambda$ is simply {{the}} distance in $\mu$m between the wavelength centers of the two FGS bands. The model does not rely on any assumed  atmospheric model as it purely uses the radiometric model to establish the precision on each target.


%

\subsection{Geometric Albedo Simulations}
To determine {{FGS's}} projected performance on measuring geometric albedos, we simulate eclipses at optical to near-IR wavelengths, accounting for both reflected light (proportional to geometric albedo) and thermal emission (a function of dayside brightness temperature). These simulations assume that a planet's albedo is a function of its temperature \citep{sudarsky00} and takes into account heat transport using the analytical expressions presented in \citet{cowan11}, \citet{schwartzcowan15}, and \citet{komacek17}. We first calculate each planet's theoretical maximum and minimum temperatures and day-to-night temperature contrast $A_{obs}$ using their calculated equilibrium temperatures \citep{komacek17}. We then estimate each planet's phase curve offset using the relationship between a planet's hot spot offset and its $A_{obs}$ value \citep{cowan11}. Assuming that the planet's longitudinal temperature structure can be represented as a combination of sine and cosine waves \citep{cowan08}, we calculate the planet's disc-averaged dayside and nightside temperature, assuming zero Bond albedo. These temperature values are then used to calculate the planet's day-to-night heat transport efficiency $\epsilon$ \citep{cowan11, schwartzcowan15}. We also estimate each planet's geometric albedo $A_{G}$ using the relationship presented in \citet{sudarsky00}, which is motivated by probable reflective condensates as a function of the equilibrium temperature. We then calculate the planet's disc-averaged dayside and nightside temperature, taking into account the planet's geometric albedo $A_{G}$ and resultant heat redistribution efficiency $\epsilon$ \citep{schwartzcowan15}. Thus we estimate the planet's eclipse depth as a function of wavelength:
\begin{align}
\delta(\lambda) = \frac{(F_{p}(\lambda))R_{p}^{2}}{(F_{s}(\lambda))R_{s}^{2}}+A_{G}\frac{R_{p}^{2}}{a^{2}}
\label{eqn:emission_with_albedo}
\end{align}
where $a$ is the semi-major axis of the planet. Assuming that to first order the planet and star both emit light according to a blackbody ($F_{p}=B(\lambda,T_{day})$ and $F_{s}=B(\lambda,T_{s}$)), and taking into account the additional reflected light due to $A_{G}$, we simulate, with uncertainties generated from the {{CASE Team's}} radiometric model, the eclipse depths $\delta(\lambda)$ and thus the visible-to-IR emission spectra observed with ARIEL/{{FGS}}+AIRS, assuming the visible ($\le$2~$\mu$m) has contributions from both thermal emission and geometric albedo $A_{G}$ while the ($>$2~$\mu$m) IR is dominated by thermal emission (i.e., $A_{G}\rightarrow0$). Thus the planet's temperature, measured by AIRS in the IR, is used to help calculate the planet's geometric albedo measured with {{FGS}}. These simulated observations are then fit by Equation~\ref{eqn:emission_with_albedo} with a Levenberg-Marquardt least-squares fitter \citep[\texttt{lmfit} Python package;][]{newville14} wrapped within a 1000-iteration Monte Carlo to conservatively account for all parameter uncertainties for both the exoplanet and its host star (planet and star radii, stellar temperature, and semi-major axis), as listed on the NASA Exoplanet Archive \citep{akeson13}.

\section{Results}
\subsection{Exoplanetary Metallicity}
We consider a mass-metallicity relationship characterized by four parameters: the slope and intercept of the main trend, the dispersion about this trend, and the pivot point where the relationship flattens for smaller-mass planets (Fig.~\ref{fig:mass_met_4panel}). Based on our simulations, we estimate that ARIEL/{{FGS}}+AIRS could measure the mass-metallicity relationship with a slope precision of 0.09~dex/M$_{\earth}$, intercept precision of 0.08~dex, dispersion precision of 0.04~dex, and pivot precision of 1.7~M$_{\earth}$ (Fig.~\ref{fig:mass_met_4panel}). These four measurements constrain the physical processes operating as the planets formed; of greatest interest is the slope---the fundamental, and currently not well constrained, relationship between planet mass and metallicity \citep{fortney13, thorngren18}. {{FGS's}} slope precision would allow it to measure the Solar System slope \citep[-1.1;][]{kreidberg14b} to 12.2$\sigma$. {{FGS's}} intercept precision would meanwhile allow it to distinguish between large (100~km) and small (1~km) planetesimals, which are predicted to differ by 0.6~dex \citep{fortney13}, to 7.5$\sigma$, and the required precision on the dispersion, will enable {{FGS}} to detect the expected 0.5~dex intrinsic variation in the relation to 12.5$\sigma$. Lastly, {{FGS's}} pivot precision would allow it to determine if Neptune-mass planets (10--20~M$_{\earth}$) are the high-mass end of the small planet distribution or the low-mass end of the giant planet population.

{{Our simulations suggest that ARIEL cannot significantly distinguish between a linear mass-metallicity relationship suggested by the Solar System’s giant planets and an assumed ``broken’’ mass-metallicity relationship (green and black models, respectively, in Fig.~\ref{fig:mass_met_4panel}) from its Type~1 population alone (assuming 1000 planets with 1 visit each). However, ARIEL will feature revisits in all three of its Types, providing for the opportunity for more precise planetary metallicties. The number of revisits for each target necessary to distinguish between these two mass-metallicity relationships will be explored in-depth in future studies.}}


\subsection{Aerosol Scattering Slope}
Based on our simulations, {{FGS}} could measure the visible-to-near-infrared spectral slope of 200 Type~1 targets with a median precision of 15~ppm/$\mu$m (Fig.~\ref{fig:slope_uncertainty}). For a hot Jupiter with a spectral modulation of 111~ppm per scale height $H_{s}$, like HD~189733b \citep{iyer16}, {{FGS}} could measure 1~$H_{s}$  of spectral modulation over 1~$\mu$m to 7.5$\sigma$; given HD~189733b's 222~ppm of spectral modulation within the H$_{2}$O band \citep[$\sim$1.2--1.4~$\mu$m;][]{iyer16}, {{FGS}} could measure this feature to  6.6$\sigma$. For a super Neptune with a spectral modulation of 57.7~ppm per scale height, like HAT-P-11b \citep{iyer16}, {{FGS}} could measure 1~$H_{s}$/$\mu$m of spectral modulation to 3.8$\sigma$; given HAT-P-11b's spectral modulation of 127~ppm in the H$_{2}$O band \citep{iyer16}, {{FGS}} could measure this feature to 3.8$\sigma$. {{FGS's}} estimated precision could distinguish between a clear and hazy atmosphere for a planet like HD~189733b to 14.5$\sigma$ \citep{pont08, pont13}. {{FGS}} could also differentiate between 0.1~$\mu$m perovskite CaTiO$_{3}$ and 0.25~$\mu$m corundum Al$_{2}$O$_{3}$ \citep{wakeford17} for a planet with a scale height like WASP-12b \citep{iyer16} to 30$\sigma$.

\begin{figure}[!htb]
\centering
\includegraphics[width=1\columnwidth]{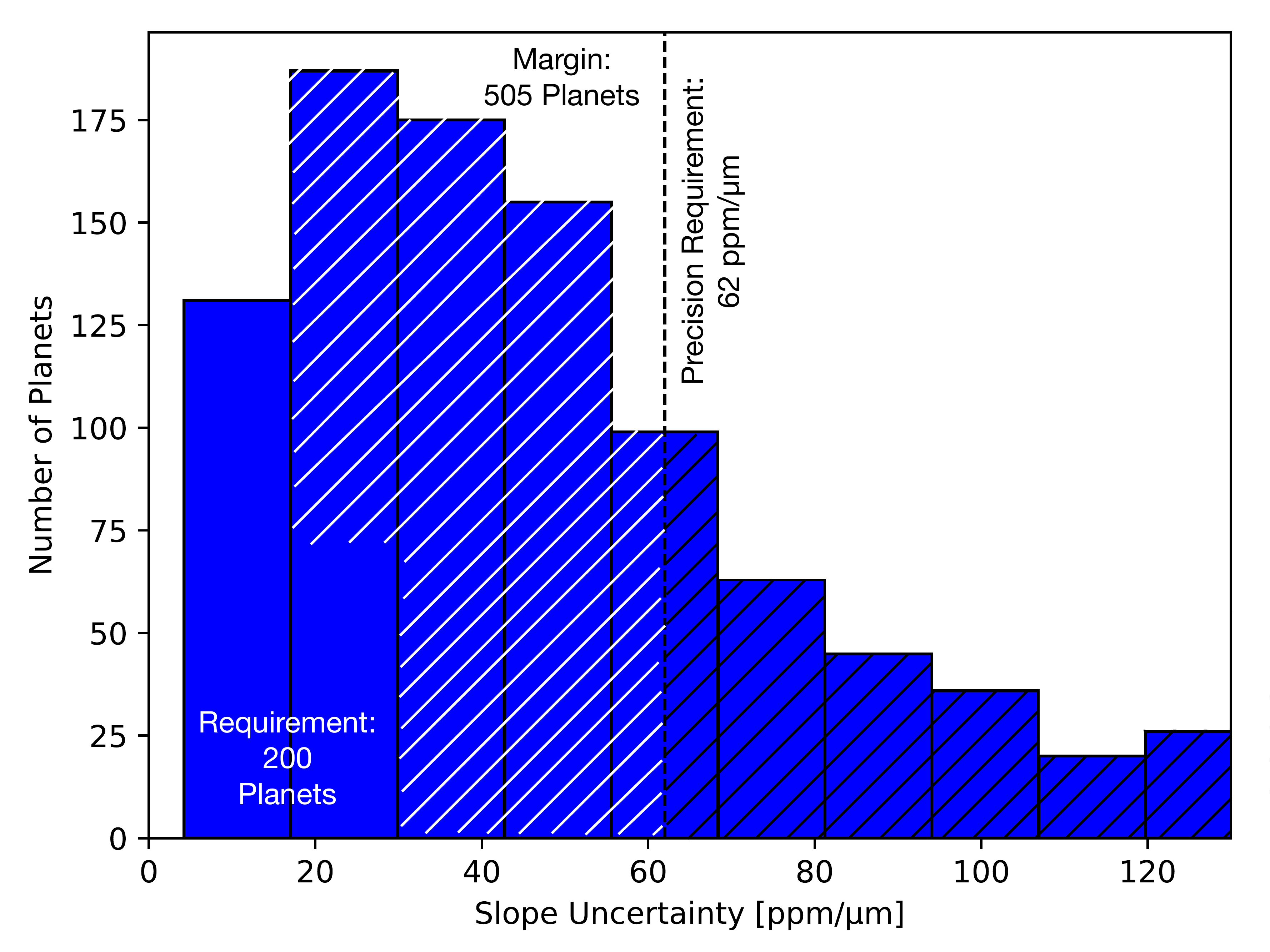}
\caption{{{{FGS}} could measure exo-atmospheric aerosol spectral slopes to high precision.} {{FGS}} could measure the median visible-to-near-infrared spectral slope of 200~Type~1 targets {{(assuming one visit of transit measurements each)}} with an uncertainty of 15~ppm/$\mu$m, enabling high precision measurements to determine if a planet's atmosphere is clear, cloudy, or hazy.
}
\label{fig:slope_uncertainty}
\end{figure}

\subsection{Geometric Albedo}
We estimate that {{FGS}} could measure the geometric albedo $A_{G}$ during eclipse for 10 Type~3 targets with a median precision of $\pm$0.02 (Fig.~\ref{fig:albedo_uncertainty}). Current geometric albedo measurements are on the order of $\lesssim$0.5 and have typical uncertainties of $\sim$0.5 \citep{rowe08, demory11, kipping11, hengdemory13, angerhausen15, esteves15, shporerhu15, bell17, dai17}. Measurements with an uncertainty of $\pm$0.1 place strong constraints on the composition of an exoplanet’s clouds \citep{parmentier16}. For example, this measurement uncertainty can determine if a planet's cloud population is Na$_{2}$S, Al$_{2}$O$_{3}$, MnS, CaTiO$_{3}$, or MgSiO$_{2}$, or is cloudless, depending on the equilibrium temperature of the planet \citep{parmentier16}. {{FGS's}} geometric albedo measurement precision could allow it to determine a planet's cloud composition to high statistical significance: for example, for a 1800~K exoplanet, {{FGS}} could distinguish between the cloud species Al$_{2}$O$_{3}$ and MgSiO$_{3}$ with a $\sim$5$\sigma$ confidence.



\begin{figure}[!htb]
\centering
\includegraphics[width=1\columnwidth]{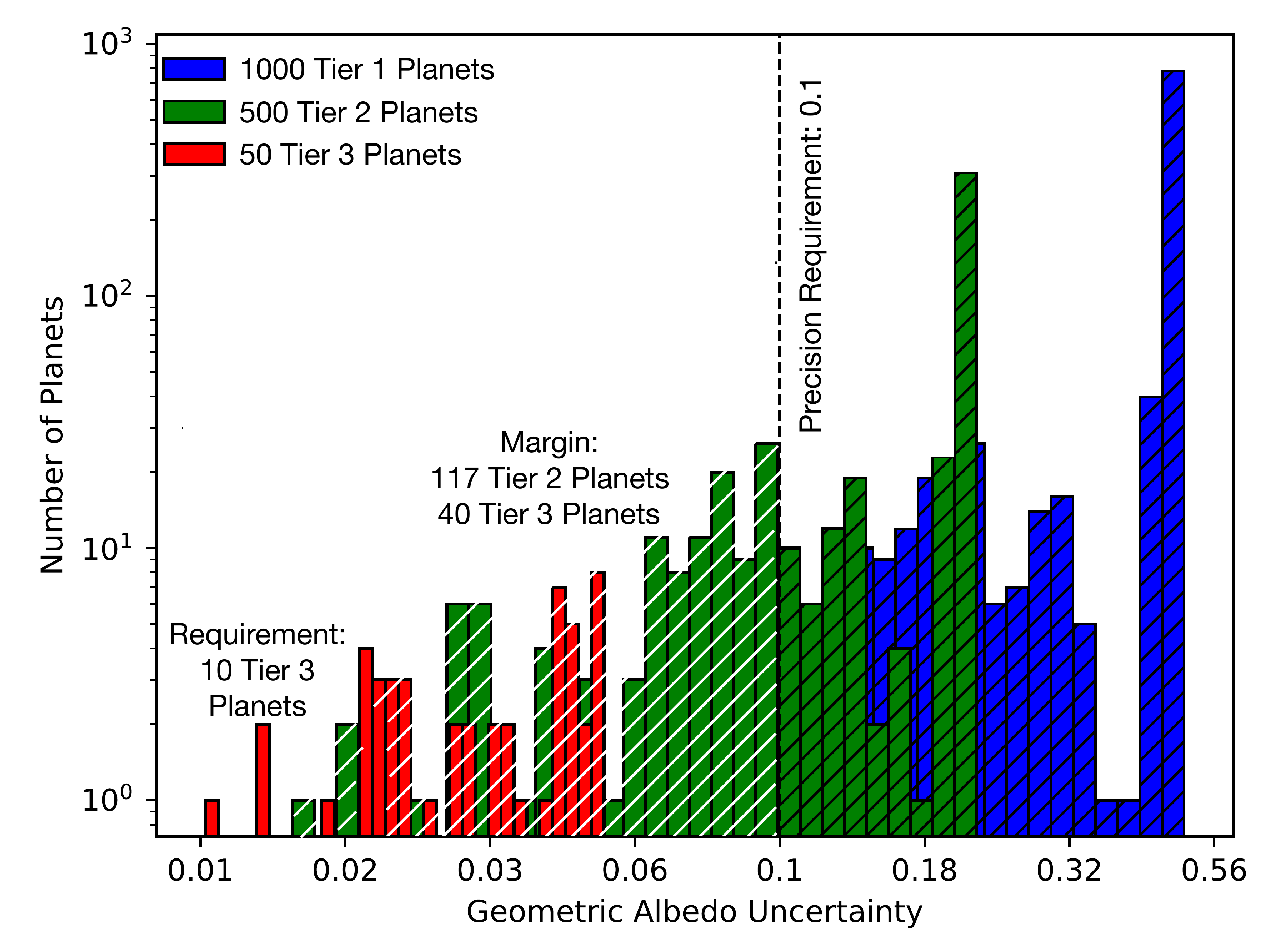}
\caption{{{{FGS}} could measure exoplanetary geometric albedos with high precision.} {{FGS}} can measure the geometric albedos of 10~Type~3 targets with a median uncertainty of 0.02.
}
\label{fig:albedo_uncertainty}
\end{figure}

\section{Conclusions}
ARIEL will survey $\sim$1000 diverse exoplanets, providing transformative science by placing the study of exoplanet atmospheres on a statistical foundation that is needed to answer some of the most important exoplanet science questions. These questions include planet formation, heat transport, and the role of composition in determining atmospheric structure. The combination of broad instantaneous spectral coverage, for a large sample of transit, eclipse, and phase curve measurements will constitute a completely unique scientific resource for the study of exoplanet atmospheres.

CASE, a proposed US Mission of Opportunity contribution to ARIEL currently in Step~2, would provide fine guidance detectors and cold front end electronics {{as a subsystem of ARIEL's FGS instrument}}, enabling visible to near-IR photometry and low-resolution spectroscopy. {{FGS}} would measure the aerosol spectral slopes and geometric albedos of the ARIEL target population. 


{{Here we presented an independent study by the CASE Team to assess some of ARIEL's science performance.}} Via science simulations we find that ARIEL can measure the mass-metallicity relationship to $>7.5\sigma$ for its 1000-planet Type~1 sample. {{Our simulations also indicate that FGS}} could achieve high precision measurements of a planet's geometric albedo and visible-to-near-infrared spectral slope: $\pm$0.02 and $\pm$15~ppm/$\mu$m on average, respectively. {{FGS}} could distinguish between clear, cloudy, and hazy skies on the order of 14.5$\sigma$. Similarly, {{FGS}} could determine a planet's aerosol composition at the 5 to 30$\sigma$ level, depending on the condensing species. {{FGS}} could thus provide significant constraints on a planet's haze and cloud compositions. 

{{Our simulations include a 30\% degradation of ARIEL's photon-noise limited performance.}} This performance margin is motivated by current Spitzer/IRAC and Hubble/WFC3 transiting exoplanet observations which achieve 1.15$\times$ and 1.33$\times$ the photon noise, respectively. We anticipate high precision measurements to be $<$15~ppm not only from ARIEL's {{FGS}} and AIRS but also JWST and other future proposed missions, such as HabEx, and LUVOIR, and the Origins Space Telescope given the exoplanet community's ability to treat systematic errors inherent in visible and near-IR detectors \citep[e.g.,][]{charbonneau08, beaulieu10, ballard10, cowan12, knutson12, deming13, lewis13, maxted13, wakeford13, crouzet14, fraine14, knutson14a, knutson14b, kreidberg14, zellem14, haynes15, sing15, wong15, demory16, evans16, sing16, tsiaras16, tsiaras16b, wong16, cartier17, damiano17, evans17, stevenson17, wakeford17, alam18, bruno18, dang18, kilpatrick18, nikolov18, wakeford18, zhang18}. Thus, {{FGS}} and AIRS will not only allow ARIEL to determine what establishes global planetary climate via high precision geometric albedo and scattering spectral slope measurements, but also place high-precision JWST observations of tens of planets \citep{cowan15} into a larger context.




\section*{Acknowledgments}
Part of the research was carried out at the Jet Propulsion Laboratory, California Institute of Technology, under contract with the National Aeronautics and Space Administration. Copyright 2018. All rights reserved.

We thank the JPL Exoplanet Science Initiative for partial support of this work.

{{We thank the anonymous referee for their helpful comments.}}

RTZ would like to thank Giovanna Tinetti, Enzo Pascale, and Ingo Waldmann for their helpful discussions about the ARIEL mission.

This research has made use of the NASA Exoplanet Archive, which is operated by the California Institute of Technology, under contract with the National Aeronautics and Space Administration under the Exoplanet Exploration Program.

\bibliography{references}

\end{document}